\documentclass{Interspeech2024}
\usepackage{hyperref}
\usepackage{array}
\usepackage{booktabs}
\usepackage{multicol}

\usepackage{xcolor}
\usepackage{listings}

\definecolor{codegreen}{rgb}{0,0.6,0}
\definecolor{codegray}{rgb}{0.5,0.5,0.5}
\definecolor{codepurple}{rgb}{0.58,0,0.82}
\definecolor{backcolour}{rgb}{0.95,0.95,0.92}

\lstdefinestyle{mystyle}{
    commentstyle=\color{codegreen},
    keywordstyle=\color{magenta},
    numberstyle=\tiny\color{codegray},
    stringstyle=\color{codepurple},
    basicstyle=\ttfamily\footnotesize,
    breakatwhitespace=false,         
    breaklines=true,                 
    captionpos=b,                    
    keepspaces=true,                 
    showspaces=false,                
    showstringspaces=false,
    showtabs=false,                  
    tabsize=2
}
\newcommand{\model}{\textit{uk-pods-conformer}\xspace}
\newcommand{\modelinit}{\textit{uk-pods-citrinet}\xspace}
\newcommand{\nvidiacitrinet}{\textit{uk-nvidia-citrinet}\xspace}
\newcommand{\data}{\textit{UK-PODS}\xspace}
\newcommand{\datainit}{\textit{UK-PODS-ALIGN}\xspace}

\interspeechcameraready

\title{Transcribe, Align and Segment: Creating speech datasets for low-resource languages.}

\name{Taras}{Sereda}
\email{taras.y.sereda@proton.me}


\begin{document}

\maketitle

\begin{abstract}
    
    In this work, we showcase a cost-effective method for generating training data for speech processing tasks. First, we transcribe unlabeled speech using a state-of-the-art Automatic Speech Recognition (ASR) model. Next, we align generated transcripts with the audio and apply segmentation on short utterances. Our focus is on ASR for low-resource languages, such as Ukrainian, using podcasts as a source of unlabeled speech.
    
    We release a new dataset \data that features modern conversational Ukrainian language. It contains over 50 hours of text audio-pairs as well as \model, a 121 M parameters ASR model that is trained on MCV-10 and \data and achieves 3x reduction of Word Error Rate (WER) on podcasts comparing to publically available \nvidiacitrinet while maintaining comparable WER on MCV-10 test split. Both dataset  \data \footnote{https://huggingface.co/datasets/taras-sereda/uk-pods} and ASR \model \footnote{https://huggingface.co/taras-sereda/uk-pods-conformer} are available on the hugging-face hub.
    
\end{abstract}

\section{Introduction}

Modern speech processing algorithms, such as Automatic Speech Recognition(ASR), Text-to-speech(TTS), require significant volumes of training data to achieve acceptable performance. Moreover, spoken language evolves itself, new words emerge, subsequently ASR models should be retrained to keep up with new terminology or speaking habits. Leveraging of self-supervised models makes usable unlabeled data, though for training down stream tasks availability of labeled data is instrumental. In this work, we explore automatic-pipeline for generation of datasets suitable for ASR modeling.

Ukrainian Mozilla Common Voice\cite{ardila2020common}(MCV-10) is a rather small dataset that contains 50 hours of training data. Fleurs\cite{conneau2022fleurs} is another multilingual dataset that spans 102 languages, though it only has 10 hours of training data for Ukrainian.

The goal of this work is to explore ways of automatic dataset creation from unlabeled audio that is of particular value for low-resource languages and domain adaptation needs, in domains such as legal, medical, finances.

\section{Related work}

Kürzinger et al.\cite{K_rzinger_2020} proposed to use ASR model trained with Connectionist temporal classification(CTC) loss to obtain frame-level token probabilities for corpus alignment and segmentation to extend amount of training data for German language.

Betker\cite{betker2023better} leverages automatically generated transcripts obtained from self-supervised model wav2vec2-large and builds a 49,000 hours corpus of English podcasts for training an expressive TTS model.

In \cite{bakhturina2022toolbox} a toolbox for speech-corpus creation is introduced. The proposed pipeline includes aligning of audio with text; segmenting audio into short utterances, and subsequent filtering of these segments, to eliminate misalignment. The approach is validated to be an effective way for obtaining additional training data for Spanish and Russian languages suitable for training ASR systems.

In this work, we adapt the toolbox proposed by Bakhturina et al. \cite{bakhturina2022toolbox} to support Ukrainian language. We use synthetic transcripts generated with Whisper \cite{radford2022whisper}, which achieves performance comparable to humans-level accuracy on a speech recognition task.

\section{Creation pipeline}

\subsection{Dataset collection}

Podcasts are of particular interest for speech processing tasks due to their accessibility, high audio quality, and diverse content. They typically feature dialogues between hosts and guests, covering various topics and capturing a broad spectrum of emotions, laughter, and subtle para-linguistic aspects of speech. In contrast, audiobooks are usually narrated by a single speaker.

We downloaded all available episodes produced by Radio Podil\footnote{\url{https://radiopodil.org/}}. These podcasts cover a range of topics, including movie reviews, book discussions, social issues, technology reviews, and the cultural life of modern Ukrainians. We collected 244 podcast episodes, with a total duration of 136 hours. The mean duration of a single episode is 33 minutes. All audio files were resampled to 16 kHz and saved as single-channel raw PCM waveforms.

\subsection{Transcription generation}

We used Whisper v3-large \cite{radford2022whisper}, a 1550 M parameters model, to generate the transcripts. Given that the model was trained on 690,000 hours of multilingual data, it possesses strong code-switching capabilities and high accuracy when transcribing English words or proper nouns that frequently occur in modern spoken Ukrainian. Notably, Whisper was trained with a next-token prediction objective, allowing it to perform causal modeling.

\subsection{Text preprocessing}
Transcript generated with Whisper contain punctuation and numbers written in digit form. Often, these transcripts are interspersed with English words, requiring to deal with code-switching scenario. We do the following text pre-processing: non-Cyrillic words are transliterated, punctuation is removed, numbers are normalized with \textit{num2words}\footnote{\url{https://github.com/savoirfairelinux/num2words}} python module. Resulting texts are converted to lower case.

\subsection{Alignment and segmentation}

For aligning text and audio, we used a CTC-trained model that predicts frame-level probability distributions over the model's vocabulary. Raw frame-level log-probabilities obtained from the baseline Citrinet checkpoint \nvidiacitrinet\footnote{\url{https://huggingface.co/nvidia/stt_uk_citrinet_1024_gamma_0_25}} together with transcripts are used as inputs for the aligner \cite{K_rzinger_2020}. This process aligns audio chunks to pieces of text and outputs an alignment confidence score for each segment candidate.

We retain only audio segments with alignment confidence score above \texttt{threshold=-2}. We follow the further filtering procedure described in the NeMo CTC-segmentation tool\footnote{\url{https://github.com/NVIDIA/NeMo/tree/main/tools/ctc_segmentation}}. First, all utterances are transcribed with the baseline ASR model. Next, we measure Character Error Rate (CER), Word Error Rate (WER), and CER on edges. Segments with a CER higher than 30\% and a WER higher than 75\% are discarded. To mitigate alignment inaccuracies at the edges, we further filter out audio segments with a CER higher than \verb|cer_threshold=60| on the edges (the first and last 5 characters) of each audio clip. We constrain audio clip duration to be longer than 1 second and shorter than 20 seconds.

\section{Dataset}

In this work, we operate on two versions of the podcast dataset. The first version is obtained through initial segmentation using the pre-trained Ukrainian Citrinet model. We call this dataset \datainit to  emphasize that it was used for fine-tuning the aligner model. It consists of 22,709 audio clips with a total duration of 33 hours. 

We use a combination of \datainit and MCV-10 to fine-tune the Ukrainian Citrinet model and obtain a better aligner model. Subsequently, we build a new dataset, \data, by re-aligning and re-segmenting the podcast corpus with \modelinit. \data consists of 34,231 audio clips with a total duration of 51 hours. The dataset statistics are listed in Table~\ref{tab:data_stats}.

We split both datasets on train/test with corresponding ratio 0.9/0.1. Each split is represented as a manifest file in jsonl format, where a single datapoint is mapped to dataclass described in the Listing~\ref{lst:datapoint}.

\begin{table}[th]
\caption{Datasets statistics}
\label{tab:data_stats}
\centering
\begin{tabular}{clccc}
\toprule
\multicolumn{2}{c}{\textbf{Dataset}} & \multicolumn{1}{c}{\textbf{Split}} & \multicolumn{2}{c}{\textbf{Duration, hours}} \\
\midrule
\multicolumn{2}{c}{MCV-10} & \multicolumn{1}{c}{train-all} & \multicolumn{2}{c}{50.1}\\
\multicolumn{2}{c}{MCV-10} & \multicolumn{1}{c}{train} & \multicolumn{2}{c}{13.6}\\
\multicolumn{2}{c}{MCV-10} & \multicolumn{1}{c}{test} & \multicolumn{2}{c}{9.8}\\

\midrule
\multicolumn{2}{c}{\datainit} & \multicolumn{1}{c}{train} & \multicolumn{2}{c}{30.5}\\
\multicolumn{2}{c}{\datainit} &  \multicolumn{1}{c}{test} & \multicolumn{2}{c}{3.4}\\
\midrule
\multicolumn{2}{c}{\data} & \multicolumn{1}{c}{train} & \multicolumn{2}{c}{46.0}\\
\multicolumn{2}{c}{\data} &  \multicolumn{1}{c}{test} & \multicolumn{2}{c}{5.1}\\

\end{tabular}
\end{table}

\break

\begin{lstlisting}[language=python, caption=Datapoint, label={lst:datapoint}]
@dataclasses.dataclass
class Datapoint:
    audio_filepath: str
    duration: float
    text: str
    text_no_processing: str
    text_normalized: str
    score: float
    pred_text: str
    wer: float
    cer: float
\end{lstlisting}

In the manifests, we keep raw text obtained from Whisper under \texttt{text\_no\_processing} key. Additionally, alignment confidence score WER and CER values are available for further filtering if needed. To validate the usefulness of the introduced dataset, \data we train the ASR model described in the next section without any additional filtering applied. For dataset analysis, we suggest using NeMo speech-dataset-explorer tool\footnote{\url{https://github.com/NVIDIA/NeMo/tree/stable/tools/speech_data_explorer}}

\section{Experiments}

We follow the recipes for cross-lingual domain-adaptation and transfer learning \cite{huang2020crosslanguage} and run two experiments. First experiment is aimed to improve alignment accuracy. In the second experiment, we fine-tune ASR on \data to demonstrate the effectiveness of the proposed method. For all experiments, NeMo \cite{kuchaiev2019nemo} was used due to availability of multiple pre-trained models and easiness of usage.

\subsection{Aligner fine-tuning}
First, we fine-tune the Ukrainian ASR model Citrinet\cite{majumdar2021citrinet} using a publicly available checkpoint from NVIDIA (\nvidiacitrinet). We employ this model as an aligner because Citrinet's fully convolutional architecture enables efficient parallel inference on long-form audio recordings. In contrast, transformer-like architectures require $O(n^2)$ memory with respect to sequence length, making them less suitable for aligning long-form audio recordings.

Citrinet has 141 M parameters, it's trained with CTC-loss on MEL-spectrograms to predict frame-level probability distributions over 1024 tokens. We don't change the tokenizer during the fine-tuning process. We fine tune model on MCV-10 train and \datainit for 50 epochs with initial learning rate set to $3\mathrm{e}{-5}$ using AdamW\cite{loshchilov2019decoupled} optimizer. We allow all model parameters to be updated.

Substantial improvement of WER by 2.5x on \datainit  test split with respect to \nvidiacitrinet suggests better fit on podcasts data distribution. This improvement allowed us to collect more audio clips after the re-segmentation and filtering pipeline, increasing total duration of audio clips in \data by 1.5x compared to \datainit.

We also observed that \modelinit degrades its performance on MCV-10 test split. We hypothesize that this occurs due to data imbalance, as the \datainit train split is almost 3x larger than MCV-10 train. All metrics are shown in Table \ref{tab:test_wer}

\begin{table}[th]
\caption{WER on MCV-10 test and \datainit test}
\label{tab:test_wer}
\centering
\begin{tabular}{clccc}
\toprule
\multicolumn{2}{c}{\textbf{Model}} & \textbf{Params} & \multicolumn{2}{c}{\textbf{WER}$\downarrow$} \\
\midrule
\multicolumn{2}{c|}{} & \multicolumn{1}{c|}{} & \textbf{MCV-10} & \textbf{\datainit}\\
\multicolumn{2}{l|}{\nvidiacitrinet} & \multicolumn{1}{c|}{141 M} & $0.096$ & $0.239$\\
\multicolumn{2}{l|}{\modelinit} & \multicolumn{1}{c|}{141 M} & $0.186$ & $0.095$\\
\end{tabular}
\end{table}

\subsection{ASR fine-tuning}
In next experiment for the base model we use Conformer, a 121 M parameters convolution-augmented transformer \cite{gulati2020conformer}, as the base model. This model is initially trained on the Russian language. For fine-tuning on Ukrainian, we train a new SentencePiece BPE \cite{kudo2018sentencepiece} tokenizer with a vocabulary constrained to 1024 tokens on MCV-10 and \data datasets. We allow all model parameters to be updated. The model is trained for 100 epochs.

Notably, we train Conformer on a larger MCV-10 split that includes all audio clips not part of the test or dev splits. This increases the total duration of MCV-10 data to 50 hours, achieving balance with \data in terms of the total number of hours.

For model regularization, we leverage InterCTC \cite{lee2021intermediate} loss on the outputs of the 1st and 8th encoder layers, with scaling coefficients set to 0.1 and 0.3, respectively. Intuitively, this allows the model to hierarchically improve its predictions from lower to higher layers.

\begin{table}[th]
\caption{WER on MCV-10 test and \data test}
\label{tab:test_wer_resegmented}
\centering
\begin{tabular}{clccc}
\toprule
\multicolumn{2}{c}{\textbf{Model}} & \textbf{Params} & \multicolumn{2}{c}{\textbf{WER}$\downarrow$} \\
\midrule
\multicolumn{2}{c|}{} & \multicolumn{1}{c|}{}  &    \textbf{MCV-10}   &  \textbf{\data}\\
\multicolumn{2}{l|}{\nvidiacitrinet} & \multicolumn{1}{c|}{141 M} & $0.096$ & $0.306$\\
\multicolumn{2}{l|}{\model} & \multicolumn{1}{c|}{121 M} & $0.116$ & $0.093$\\

\end{tabular}
\end{table}

The resulting model \model obtains significant improvement on \data while maintaining low WER on MCV-10, listed in Table \ref{tab:test_wer_resegmented}. We hypothesize this is achieved due to the balanced data used for model training. The model demonstrates a higher robustness to background noise and music, it accurately transcribes domain specific terminology, and foreign proper nouns. We suggest this is achieved due to the diversity of the topics covered in the podcasts, as well as various background sounds and recording environments. We validate this by inspecting transcripts for audio clips with background music, where \model consistently performs better than \nvidiacitrinet despite having only 121 M parameters.

All models were trained on a small deep learning cluster of 6 NVIDIA RTX 3090 GPUs.

\section{Conclusion}
In this work, we validated usefulness of automatically generated transcripts with subsequent corpus alignment and segmentation for dataset generation.This method allows for the domain adaptation of ASR systems. We expect this method to be general enough to apply to other low-resource languages. As future work, we plan to investigate whether the introduced dataset can be useful for other speech processing tasks, particularly for TTS.

\section{Acknowledgement} We would like to express our gratitude to Pavlo Bondarenko, producer and podcast host at Radio Podil for granting as permission to use the podcasts.

We also thank Oleksiy Syvokon and Volodymyr Kyrylov for their valuable feedback and suggestions on the early version of this paper.

\bibliographystyle{IEEEtran}
\bibliography{resources}

\end{document}